\documentclass[prd,twocolumn,showpacs,showkeys,preprintnumbers,amsmath,amssymb]{revtex4} 
\newcommand{\bv}{\bar{\varphi}} 
\newcommand{\chris}{{\Gamma}^{k}_{ij}} 
\newcommand{\intp}[1]{\int\frac{d^n{#1}}{(2\pi)^n}}
\begin{document}
\title{Quantum gravity and scalar fields}
\author{Paul T. Mackay and David J. Toms}
\homepage{http://www.staff.ncl.ac.uk/d.j.toms} \email{d.j.toms@newcastle.ac.uk}
\affiliation{School of Mathematics and Statistics, Newcastle University, Newcastle upon Tyne, United Kingdom, NE1 7RU}
\date{\today}
\begin{abstract}
In this paper we consider the quantization of a scalar field coupled to gravity at one loop order. We investigate the divergences appearing in the mass (i.e. $\phi^2$) term in the effective action. 
We use the Vilkovisky-DeWitt effective action technique which guarantees that the result is gauge invariant as well as gauge condition independent in contrast to traditional calculations. Our final result is to identify the complete pole part of the effective action.

\end{abstract}

\maketitle

In a remarkable paper Robinson and Wilczek~\cite{robinsonwilczek} performed a calculation in quantized Yang-Mills-Einstein gravity showing that the gauge coupling constant for Yang-Mills theory received a purely quantum gravity contribution to the $\beta$ function in addition to that normally present in the absence of gravity~\cite{GrossWilczek,Politzer}. This new contribution tended to result in asymptotic freedom even for theories that in the absence of gravity were not asymptotically free. The calculation used the traditional background field method that is equivalent to conventional Feynman diagram methods. However, if we are interested in a field theory which includes gravity then we encounter a problem - the seminal work of \cite{thooftveltman,deservan1,deservan2,deservan3} showed that quantum gravity is non-renormalizable. Nevertheless, advances in the understanding of effective field theory methods~\cite{Donoghue1,Donoghue2,burgesslivrev} lends credence to obtaining physical predictions for a non-renormalizable theory like gravity. As long as we are only interested in physics at a scale $E\ll M_P$, where $M_P$ is the Planck mass, then the results of the effective theory should coincide with the results predicted by the underlying fundamental theory, whatever its nature. Phenomenological consequences of the Robinson-Wilczek result were studied in \cite{gogoladze} and it has been suggested \cite{huang} that the results can be used to substantiate the weak gravity conjecture \cite{AHetal,Banks}.

Further analysis cast doubt on the results of Robinson and Wilczek~\cite{robinsonwilczek}. Pietrykowski~\cite{pietrykowski} repeated their calculation, showed that the result was gauge condition dependent, and that it was possible to obtain no quantum gravitational contribution to the renormalization group $\beta$ function. A method which is naturally gauge invariant and gauge condition independent (that we will describe below) was used along with dimensional regularization to show that a vanishing quantum gravitational contribution to the $\beta$ function was obtained for Einstein-Maxwell theory~\cite{toms1}. Ebert, Plefka and Rodigast~\cite{ebert1} used a traditional Feynman diagram approach to show that there was no purely quantum gravity contribution to the $\beta$ function in Einstein-Yang-Mills theory. Their work is especially interesting because they showed that use of a cutoff agreed with dimensional regularization, with gauge invariance retained by requiring that the Slavnov-Taylor-Ward-Takahashi identities be satisfied. Tang and Wu~\cite{tangwu} suggested an unconventional regularization scheme, namely loop regularization, which they claimed handled the quadratic divergences correctly. Their paper, in contrast to earlier work, claimed that their regularization method led to non-zero gravitational contributions. Further work \cite{ebert2} has also been done in dimensions higher than 4 which looked at the ADD scenario in extra-dimensional braneworlds, and in studying Lee-Wick terms \cite{wuzhong1,wuzhong2,rodigastschuster1}.

All of the calculations related to the Robinson and Wilczek result described so far were for pure Einstein gravity. In \cite{toms2,toms3} Einstein gravity with a cosmological constant was considered and it was shown that the cosmological constant could change the behaviour of the running from that of pure gravity. Although the result was not the same as that found by Robinson and Wilczek~\cite{robinsonwilczek} at least in Einstein-Maxwell theory the consequences were similar - quantum gravity tends to lead to asymptotic freedom. The purpose of the present paper is to consider the mass and field renormalizations for a quantized scalar field in the presence of gravity with a cosmological constant using the methods of \cite{toms1,toms2,toms3} and relate this to previous work. Recently Rodigast and Schuster~\cite{rodigastschuster2} looked at the gravitational contributions to $\phi^4$ and Yukawa couplings and concluded that the gravitational contributions to the running of the masses vanish. \cite{zanusso} also examined Yukawa couplings using a completely different method. It was shown some time ago \cite{antoniadis1,witdass} that mass counterterms in quantum gravity when computed using standard methods depend on the gauge condition. We will look at the gauge condition dependence of the mass renormalization again using the background field method of Vilkovisky~\cite{Vilkovisky1,Vilkovisky2} and DeWitt~\cite{DeWittVD} that in a natural way leads to a gauge condition independent result. (See \cite{parkertoms} for a comprehensive review.) By keeping our calculation sufficiently general we will show explicitly the gauge condition dependence that is found in traditional diagrammatic approaches.

The Vilkovisky-DeWitt effective action to one loop order is (using DeWitt notation \cite{dewitt2} where the index $i$ contains the coordinate and all other labels of the field)
\begin{align}\label{VDmain}
\Gamma[\bv] &= S[\bv]-\ln\det Q_{\alpha\beta}[\bv] \nonumber \\
&+\frac{1}{2}\lim_{\alpha \to 0} \ln \det \left( \nabla^i\nabla_j S[\bv]+\frac{1}{2\alpha}K^i_\alpha[\bv]K^\alpha_j[\bv]\right).
\end{align}
Here $S[\bv]$ is the classical action, $Q_{\alpha\beta}$ is the ghost term,  and $K^i_\alpha$ are the generators of gauge transformations. We will be brief here as the basic results have been repeated already in \cite{toms1,toms2,toms3}. The covariant derivative is
\begin{equation}\label{covarderiv}
\nabla_i\nabla_jS[\bv]=S_{,ij}[\bv]-\bar{\Gamma}^k_{ij}S_{,k}[\bv]
\end{equation}
where $\bar{\Gamma}^k_{ij}$ is the connection term that is crucial for obtaining a gauge condition independent result.

The connection term in (\ref{covarderiv}) will vanish if $S_{,i}=0$, {\em i.e.\/} when the background field is a solution to the classical equations of motion. So expanding about a Minkowksi metric, which is not a classical solution in general if matter fields or a cosmological constant are present, requires the connection term for gauge condition independence. This will be demonstrated explicitly below.

We can replace the connection $\bar{\Gamma}^k_{ij}$ in (\ref{covarderiv}) by the standard Christoffel symbol $\chris$ formed from the metric on the space of fields if we choose the Landau-DeWitt gauge condition
\begin{equation}\label{LandauDeWitt}
\chi_\alpha=K_{\alpha i}[\bv]\eta^i=0,
\end{equation}
where we have split the fields into a background $\bv^i$ and quantum part $\eta^i$ as
\begin{equation}\label{bfmethod}
\varphi^i=\bv^i+\eta^i.
\end{equation}
It can be shown~\cite{FradkinTseytlin,Rebhan,parkertoms} that the Landau-DeWitt gauge coincides precisely with the gauge condition independent result with any other gauge choice.

We are interested here in a scalar field coupled to gravity. We choose the action
\begin{equation}\label{VD3.1}
S=S_M+S_G,
\end{equation}
where
\begin{align}\label{VD3.2} S_M=&\int d^nx|g(x)|^{1/2}
\Big\lbrace \frac{1}{2}\partial^\mu\varphi\partial_\mu\varphi+\frac{1}{2}m^2\varphi^2\nonumber \\
&+\frac{1}{2}\xi R\varphi^2+\frac{\lambda}{4!}\phi^4\Big\rbrace
\end{align}
describes a massive scalar field, with a $\phi^4$ interaction and some non-minimal coupling to the curvature measured by the parameter $\xi$, and
\begin{equation}\label{VD3.3}
S_G=-\frac{2}{\kappa^2}\int d^nx |g(x)|^{1/2} (R-2\Lambda),
\end{equation}
is the gravitational Einstein-Hilbert action with the inclusion of a cosmological constant $\Lambda$ where $\kappa^2=32\pi G$ in terms of the gravitational constant $G$.

At this stage we split the field into a background part and a quantum part. Since we are using a gauge invariant method we are free to choose whichever background metric we desire, not necessarily a solution to the classical field equations. We choose a flat metric tensor for the gravitational part, adopt a Euclidean (rather than a Minkowskian) signature for convenience, and keep the background scalar general at this stage. We have
\begin{eqnarray}
g_{\mu\nu}&=&\delta_{\mu\nu}+\kappa h_{\mu\nu},\label{hmunu}\\
\varphi(x)&=&\bar{\varphi}(x)+\psi(x).\label{scalars}
\end{eqnarray}

We can write the third term in (\ref{VDmain}) as
\begin{align}\label{GammaG}
\Gamma_G&=\frac{1}{2}\ln\det\left\lbrace\nabla^i\nabla_j S[\bv]+\frac{1}{2\alpha}K^i_\alpha[\bv]K^\alpha_j[\bv] \right\rbrace \nonumber \\
&=-\ln\int[d\eta]e^{-S_q}
\end{align}
where
\begin{equation}\label{Sq}
S_q=\eta^i\eta^j\left(\nabla_i\nabla_jS+\frac{1}{2\alpha}K_i^\alpha K_{j\alpha}\right).
\end{equation}
The limit $\alpha\rightarrow0$ is understood here. We then proceed to expand in powers of the interaction up to quadratic order (since we are interested in the mass and field renormalization terms). We will write
\begin{equation}\label{Sqexpanded}
S_q=S_0+S_1+S_2
\end{equation}
with the subscripts counting the order of the background fields. There are also cubic and quartic terms in the background scalar field, but these cannot contribute to the mass or field renormalizations.

It remains to calculate the three terms of (\ref{Sq}) where the covariant derivatives are expanded in terms of ordinary ones and the connection using (\ref{covarderiv}). The first term, containing two ordinary derivatives of the action, follows from (\ref{VD3.1}) using (\ref{hmunu}) and (\ref{scalars}) and retaining only those terms quadratic in the quantum fields. For the connection term, the factor containing one derivative can similarly be read off from the linear term in (\ref{VD3.1}). The connection itself results from considering a metric on the space of fields. Its non-zero components are the diagonal terms, the scalar part
\begin{equation}
g_{\varphi(x)\varphi(x')}=\sqrt{g(x)}\delta(x,x'),
\end{equation}
and the gravitational component
\begin{align}
g_{g_{\mu\nu}(x)g_{\rho\sigma}(x')}&=\sqrt{g(x)}\left\{ g^{\mu(\rho}(x)g^{\sigma)\nu}(x)\right. \\
&\left.- \frac{1}{2}g^{\mu\nu}(x)g^{\rho\sigma}(x) \right\}\delta(x,x')
\end{align}
where the brackets around the indices denote a symmetrization of the form $T_{(\mu\nu)}=\frac{1}{2}\left(T_{\mu\nu}+T_{\nu\mu}\right)$. It is now easy to evaluate the Christoffel connection.

Finally, we calculate the gauge fixing term.  We wish to deduce the generators $K^i_\alpha$ of the infinitesimal gauge transformations written in condensed notation as
\begin{equation}
\delta\varphi^i=K^i_\alpha[\varphi]\delta\epsilon^\alpha
\end{equation}
for some infinitesimal parameters $\delta\epsilon^\alpha$. Here the infinitesimal parameters represent infinitesimal coordinate changes. For the scalar field part, it is easy to show that an infinitesimal change in the field is given by
\begin{equation}\label{changescalar}
\delta\varphi=-\delta\epsilon^\mu\partial_\mu\varphi
\end{equation}
whilst for gravity a consideration of the metric transformation under an infinitesimal change of coordinates gives
\begin{equation}\label{changegravity}
\delta g_{\mu\nu} = -\delta\epsilon^\lambda g_{\mu\nu,\lambda}-\delta\epsilon^\lambda \,_{,\mu}g_{\lambda\nu}-\delta\epsilon^\lambda \,_{,\nu}g_{\lambda\mu}.
\end{equation}
Using the condition (\ref{LandauDeWitt}) with (\ref{changescalar}) and (\ref{changegravity}) gives
\begin{equation}
\chi_{\lambda}(x)=\frac{2}{\kappa}(\partial^\mu h_{\mu\lambda}-\frac{1}{2}\partial_\lambda h ) -\omega \partial_\lambda\bar{\varphi}\psi
\end{equation}
with a parameter $\omega$ introduced to highlight the gauge condition dependence of the conventional method. It must be set to unity to obtain the gauge condition independent result. The generators of gauge transformations are easily read off from (\ref{changescalar}) and (\ref{changegravity}).

Putting these results together and separating the terms by their order in the background fields, as in (\ref{Sqexpanded}), we find
\begin{eqnarray}
S_0&=&\int d^nx\left\lbrace-\frac{1}{2}h^{\mu\nu}\Box h_{\mu\nu}+\frac{1}{4}h\Box h\right.\nonumber \\
&&\left.\qquad+\left(\frac{1}{\alpha}-1\right)\left(\partial^\mu h_{\mu\nu}-\frac{1}{2}\partial_\nu h \right)^2\right.\nonumber\\
&&\left.\qquad-\Lambda\left(h^{\mu\nu}h_{\mu\nu} -\frac{1}{2}h^2\right) \left\lbrack 1+\frac{v}{2}\left(\frac{n-4}{2-n}\right) \right\rbrack\right.\nonumber\\
&&\qquad\left.+\frac{1}{2}\partial^\mu\psi\partial_\mu\psi+\frac{1}{2}m^2\psi^2 +\frac{vn\Lambda}{4-2n}\psi^2 \right\rbrace,\label{S0}\\
S_1&=&\kappa\int d^nx     \left\lbrace \frac{1}{2}(h\delta^{\mu\nu}-2h^{\mu\nu})\partial_\mu\bv\partial_\nu\psi +\frac{1}{2}m^2\bv h\psi \right.\nonumber\\
&&+\xi\bv (h^{\mu\nu}{}_{,\mu\nu}-\Box h)\psi  -\frac{\omega}{\alpha}(\partial^\mu h_{\mu\nu} -\frac{1}{2}\partial_\nu h)\partial^\nu\bv\psi \nonumber \\
&&\left.-\frac{v}{4}(-\Box\bv+m^2\bv)h\psi
\right\rbrace,\label{S1}\\
S_2&=&\kappa^2\int d^nx     \left\lbrace \frac{1}{2}\left(h^{\mu\lambda}h_{\lambda}{}^{\nu} -\frac{1}{2}hh^{\mu\nu} -\frac{1}{4}\delta^{\mu\nu}h^{\alpha\beta}h_{\alpha\beta} \right.\right.\nonumber \\
&&\left.\left.\qquad+\frac{1}{8}h^2\delta^{\mu\nu}\right)\partial_\mu\bv\partial_\nu\bv
\right.\nonumber\\
&&\qquad+\frac{1}{2}\left(\frac{1}{8}h^2-\frac{1}{4}h^{\mu\nu}h_{\mu\nu}\right)m^2\bv^2 \nonumber \\
&&\qquad+\frac{1}{2}\xi(R_2+\frac{1}{2}hR_1)\bv^2+\frac{\lambda}{4\kappa^2}\bv^2\psi^2\label{S2}\\
&&\qquad+\frac{v}{4}h_{\mu\nu}h_{\lambda\sigma}\left\lbrack\frac{1}{2}\delta^{\mu\nu}T^{\lambda\sigma} -\delta^{\mu\lambda}T^{\nu\sigma}\right.\nonumber \\
&&\left.\qquad+\frac{1}{4(n-2)}T\left( \delta^{\mu\sigma}\delta^{\nu\lambda} +\delta^{\mu\lambda}\delta^{\nu\sigma}-\delta^{\mu\nu}\delta^{\lambda\sigma}\right)\right\rbrack \nonumber\\
&&\left.\qquad-\frac{v}{8(2-n)}T\psi^2
+\frac{\omega^2}{4\alpha}(\partial^\mu\bv\partial_\mu\bv)\psi^2\right\rbrace,
\end{eqnarray}
where $R_2+\frac{1}{2}hR_1$ is the quadratic part of $|g|^{1/2}R$ given by
\begin{eqnarray}
R_2+\frac{1}{2}hR_1&=&h^{\mu\nu}\Box h_{\mu\nu} -2h_{\mu\nu}\partial^{\mu}\partial^{\lambda}h_{\lambda}{}^{\nu} -\partial^\lambda h_{\lambda}{}^{\mu} \partial^\nu h_{\mu\nu} \nonumber\\
&&\hspace{-2cm}+\partial^\lambda h_{\lambda}{}^{\mu}\partial_\mu h +h^{\mu\nu}\partial_\mu\partial_\nu h +\frac{3}{4}\partial^\lambda h^{\mu\nu}\partial_\lambda h_{\mu\nu}-\frac{1}{4}\partial^\lambda h\partial_\lambda h \nonumber\\
&&\hspace{-2cm} -\frac{1}{2}\partial^\lambda h^{\mu\nu}\partial_\mu h_{\lambda\nu} +\frac{1}{2}h\partial_\mu\partial_\nu h^{\mu\nu} -\frac{1}{2}h\Box h.
\end{eqnarray}
We have defined $T_{\mu\nu}$ to represent the energy-momentum tensor terms of order $\bv^2$ given by
\begin{eqnarray}
T_{\mu\nu}&=&\partial_\mu\bv\partial_\nu\bv-\frac{1}{2}\delta_{\mu\nu} \partial^\alpha\bv\partial_\alpha\bv -\frac{1}{2}\delta_{\mu\nu}m^2\bv^2\nonumber \\
&&+\xi\delta_{\mu\nu}(\Box\bv^2) -\xi\partial_\mu\partial_\nu\bv^2,\\
T=&&\hspace{-0.5cm}\left(1-\frac{n}{2}\right)\partial^\mu\bv\partial_\mu\bv-\frac{n}{2}m^2\bv^2 +(n-1)\xi\Box\bv^2.
\end{eqnarray}
The parameter $v$ has been introduced to indicate the terms that arise from the connection terms in the covariant derivative. For the gauge condition independent result we must take $v=1$. However by taking $v=0$ we can obtain the result of using standard methods, and this will allow us to illustrate the gauge condition dependence of the traditional result explicitly.

We can now evaluate the effective action to quadratic order in the background scalar field. Expanding (\ref{GammaG}) gives us
\begin{equation}\label{Wick}
\Gamma_G=\langle S_2\rangle-\frac{1}{2}\langle S_1^2\rangle
\end{equation}
where Wick's theorem is used to compute the expressions in angular brackets. We use the basic pairings
\begin{eqnarray}
\langle \psi(x)\psi(x')\rangle&=&G(x,x'),\label{VD4.18}\\
\langle h_{\rho\sigma}(x)h_{\lambda\tau}(x')\rangle&=&G_{\rho\sigma\lambda\tau}(x,x'),\label{VD4.19}
\end{eqnarray}
where the massive scalar propagator is
\begin{equation}
G(x,x')=\intp{p}e^{ip\cdot(x-x')}G(p)
\end{equation}
and the graviton propagator is
\begin{equation}
G_{\alpha\beta\mu\nu}(x,x') = \intp{p} e^{ip\cdot(x-x')}G_{\alpha\beta\mu\nu}(p).
\end{equation}
From $S_0$, we find
\begin{equation}
G(p)=\frac{1}{p^2+M^2}
\end{equation}
with
\begin{equation}
M^2=m^2+\frac{nv\Lambda}{2-n}
\end{equation}
and
\begin{align}
G_{\alpha\beta\mu\nu}(p) &=\frac{\delta_{\alpha\mu}\delta_{\beta\nu}+\delta_{\alpha\nu}\delta_{\beta\mu} -\frac{2}{n-2}\delta_{\alpha\beta}\delta_{\mu\nu}}{2\left(p^2-2\lambda\right)} \nonumber \\
&\hspace{-1.5cm}+\frac{1}{2}(\alpha-1)\frac{\delta_{\alpha\mu}p_\beta p_\nu+\delta_{\alpha\nu}p_\beta p_\mu+\delta_{\beta\mu}p_\alpha p_\nu+\delta_{\beta\nu}p_\alpha p_\mu}{\left(p^2-2\lambda\right) \left(p^2-2\alpha\lambda\right)}
\end{align}
where
\begin{equation}
\lambda=\Lambda+v\Lambda\left(\frac{n-4}{4-2n}\right)
\end{equation}
can be set equal to $\Lambda$ in the limit $n \to 4$.
Using the propagator pairings leads to lengthy expressions involving products of propagators. We utilize dimensional regularization~\cite{tHooft3} and let the dimension $n \to 4$.

After considerable calculation, we find the gauge part of the effective action in (\ref{Wick}) to be of the form (keeping only terms quadratic in the background scalar field)
\begin{equation}
\Gamma_G=\kappa^2L\int d^4x \left\lbrace A(\Box\bv)^2+B\bv\Box\bv+C\bv^2\right\rbrace \label{GammaGresult}
\end{equation}
where $L=-\frac{1}{8\pi^2(n-4)}$ contains the divergent pole part as $n \to 4$ and
\begin{eqnarray}
A&=&\frac{3}{16}v^2-\frac{1}{8}\omega v-\frac{1}{16}\alpha v^2-\frac{1}{4}\xi-\frac{3}{8}v-\frac{1}{4}\omega +\frac{1}{4}\alpha v,\nonumber \\
B&=&\frac{1}{8}\Lambda {v}^{2} + \frac{17}{16} {m}^{2} v + \frac{1}{8} \omega {m}^{2} v + \frac{1}{4} \omega {m}^{2} - \frac{3}{8} {m}^{2} {v}^{2}\nonumber\\
&& - \frac{1}{2} \alpha {m}^{2} v + \frac{1}{8} \alpha {m}^{2} {v}^{2} - \frac{3}{4} {m}^{2} + \frac{3}{4}{\xi}^{2} {m}^{2}\nonumber\\
&& - \frac{3}{2} \Lambda {\xi}^{2} v + \frac{1}{4} \alpha {m}^{2} - \frac{3}{4} \xi {m}^{2} v + \frac{3}{2} \Lambda \xi v^2 + \frac{3}{4} \xi {m}^{2}\nonumber\\
&& - \frac{3}{2} \Lambda \xi - \Lambda \omega v + \frac{1}{2} \Lambda \alpha v + \frac{1}{2} \Lambda {\omega}^{2} - \Lambda \alpha \omega%
 + \frac{1}{2} \Lambda {\alpha}^{2},\nonumber\\
C&=&-\frac{3}{2}\Lambda m^2 -\Lambda\alpha^2 m^2-3\xi\Lambda^2+\frac{\lambda v\Lambda}{2\kappa^2} -\frac{\lambda m^2}{4\kappa^2} \nonumber \\ 
&&-\frac{1}{4}\Lambda m^2 v^2 -\frac{5}{8}m^4v +\frac{3}{4}m^4 +\frac{3}{16}m^4v^2-\frac{1}{4}\alpha m^4\nonumber\\
&&+\frac{1}{4}\alpha m^4v-\frac{1}{16}\alpha m^4v^2 +3\Lambda\xi m^2+\frac{3}{2}\Lambda\xi m^2v \nonumber \\
&&-\frac{3}{2}\Lambda\xi m^2v^2-\frac{3}{2}\xi m^4+\frac{3}{4}\xi m^4v+3\Lambda^2\xi^2-\frac{3}{2}\Lambda\xi^2 m^2 \nonumber \\ &&+3\Lambda^2\xi^2 v +\frac{3}{4}\xi^2 m^4 -3\Lambda\xi^2 m^2v+3\Lambda^2\xi^2v^2.\label{ABC}
\end{eqnarray}
Notice that there are no terms singular as $\alpha \to 0$ as required from (\ref{VDmain}). Although not shown here, such singular terms do appear separately in the expressions for $\langle S_2\rangle$ and $\langle S_1^2\rangle$ at the intermediate steps of the calculation. There is still the ghost term to contend with; however, it is easy to show that it only makes a pole contribution of quartic order and therefore (\ref{GammaGresult}) is the total divergent part of the effective action at quadratic order.

The traditional, gauge condition dependent result comes from setting $v=0$. It can be seen that the results of the traditional method depend on the gauge parameters $\omega$ and $\alpha$ even when we set $v=0$. The correct gauge condition independent  result is found by setting $v=\omega=1$, and also taking the limit $\alpha \to 0$. We find the form of (\ref{GammaGresult}) with
\begin{eqnarray}
A&=& -\frac{9}{16}-\frac{1}{4}\xi,\\
B&=&\left (\frac{5}{16}+\frac{3}{4}\xi^2\right)m^2-\left(\frac{1}{2}+\frac{3}{2}\xi^2\right)\Lambda,\\
C&=& -\frac{7}{4}\Lambda m^2 - 3\xi\Lambda^2+\frac{\lambda\Lambda}{2\kappa^2}-\frac{\lambda m^2}{4\kappa^2}+\frac{5}{16}m^4 +3\Lambda\xi m^2\nonumber \\
&&-\frac{3}{4}\xi m^4 + 9\Lambda^2\xi^2-\frac{9}{2}\Lambda\xi^2 m^2 +\frac{3}{4}\xi^2m^4.
\end{eqnarray}
For the case of pure gravity (no cosmological constant) with a minimally coupled scalar field ($\xi=0$) we find
\begin{eqnarray}
A&=& -\frac{9}{16},\\
B&=& \frac{5}{16}m^2,\\
C&=& -\frac{\lambda m^2}{4\kappa^2}+\frac{5}{16}m^4.
\end{eqnarray}

To renormalize the scalar field theory we write the bare field and mass as \cite{tHooftRG}
\begin{eqnarray}
\bv_{\rm Bare}&=&\mu^{n/2-2}Z_{\varphi}^{1/2}\bv,\\
m_{\rm Bare}^2&=&m^2+\delta m^2,
\end{eqnarray}
with $\mu$ the unit of mass. The counterterm part of the classical action (\ref{VD3.2}), for $g_{\mu\nu}=\delta_{\mu\nu}$, becomes
\begin{equation}
\delta S_M=\int d^4x\left\lbrace -\frac{1}{2}Z_{\varphi}\bv\Box\bv +\frac{1}{2}(\delta m^2+m^2\delta Z_\varphi)\bv^2\right\rbrace
\end{equation}
to quadratic order in $\bv$ with $Z_\varphi=1+\delta Z_\varphi$. These counterterms must absorb the relevant poles of the effective action (\ref{GammaGresult}), so we find
\begin{eqnarray}
\delta Z_\varphi&=&-\frac{\kappa^2 B}{4\pi^2(n-4)},\\
\delta m^2&=&\frac{\kappa^2 (C+m^2B)}{4\pi^2(n-4)}.
\end{eqnarray}
As a check on our results, in the absence of gravity (taking $\kappa^2\rightarrow0$ and $\Lambda\rightarrow0$) we find $\delta Z_\varphi=0$ and 
\begin{equation}
\delta m^2=-\frac{\lambda m^2}{16\pi^2(n-4)}
\end{equation}
in agreement with standard flat spacetime results. (See Collins~\cite{Collins} for example.) Even for $\Lambda=\xi=0$ (but $\kappa\ne0$) we find $\delta m^2\ne0$ in contradiction to the result of \cite{rodigastschuster2} where it was claimed that $\delta m^2=0$. Note that if we set $\Lambda=\xi=v=\omega=0$ and $\alpha=1$ in (\ref{ABC}), then our results for $A,B,C$ confirm the calculations of \cite{rodigastschuster2}.

A subsequent paper \cite{mackaytoms} will contain more details of the above calculation, a complete analysis of the pole part of all of the derivative terms, and of the $\phi^4$ terms and the running of its coupling parameter. All of these results will be independent of gauge condition. 

We made use of the programs FORM~\cite{form} and Cadabra~\cite{cadabra3} for some of the lengthy 
manipulations.

One of us (PTM) would like to acknowledge EPSRC for funding this research. We are grateful to A. Rodigast and T. Schuster for questioning a previous version of the results that led to a correction for $B$.


\end{document}